# Evaluation of 3D gold nanodendrite layers obtained by templated galvanic displacement reactions for SERS sensing and heterogeneous catalysis[†]

Weijia Han,[a] Elzbieta Stepula,[b] Michael Philippi,[a] Sebastian Schlücker[b] and Martin Steinhart[a]*

[a] Institut für Chemie neuer Materialien, Universität Osnabrück, Barbarastr. 7, 49076 Osnabrück, Germany

[b] Physical Chemistry I, Department of Chemistry and Center for Nanointegration Duisburg-Essen (CENIDE), University of Duisburg-Essen, Universitätsstraße 5, 45141 Essen, Germany

[†] Electronic supplementary information (ESI) available. See DOI:

**Abstract**

Dense layers of overlapping three-dimensional (3D) gold nanodendrites characterized by high specific surfaces as well as by abundance of sharp edges and vertices creating high densities of SERS hotspots are promising substrates for SERS-based sensing and catalysis. We have evaluated to what extent structural features of 3D gold nanodendrite layers can be optimized by the initiation of 3D gold nanodendrite growth at gold particles rationally positioned on silicon wafers. For this purpose, galvanic displacement reactions yielding 3D gold nanodendrites were guided by hexagonal arrays of parent gold particles with a lattice constant of 1.5 µm obtained by solid-state dewetting of gold on topographically patterned silicon wafers. Initiation of the growth of dendritic features at edges of the gold particles resulted in the formation of 3D gold nanodendrites while limitation of dendritic growth to the substrate plane was prevented. The regular arrangement of the parent gold particles supported the formation of dense layers of overlapping 3D gold nanodendrites that were sufficiently homogeneous within the resolution limits of Raman microscopes. Consequently, SERS mapping experiments revealed a reasonable degree of uniformity. The proposed preparation algorithm comprises only bottom-up process steps that can be carried out without use of costly instrumentation.

Dendritic precious metal nanostructures show excellent catalytic properties owing to their structural features and their large specific surfaces, as evidenced by model reactions such as the electrocatalytic oxidation of methanol [1,2] and ethanol [3] as well as the reduction of 4-nitrophenol.[4-6] Furthermore, precious metal nanodendrites possess morphological features like sharp edges and vertices as well as nanoscale junctions that enable highly efficient preconcentration sensing by surface-enhanced Raman scattering (SERS).[1, 4, 7-13] SERS is related to the enhancement of electromagnetic fields caused by surface plasmons of metal substrates exhibiting plasmon resonances in the visible spectral range.[14-17] SERS enables trace analysis as well as *in situ* monitoring of molecular processes because the Raman scattering intensity originating from analytes attached to SERS-active metals and, consequently, the corresponding detection sensitivities are drastically increased. Strong local electromagnetic field enhancement accompanied by strong SERS occurs, in particular, at edges and vertices of metallic nanoobjects.[18-20] As compared to other types of precious metal nanostructures, nanodendrites show excellent SERS activities[21] and catalytic performance[22] because they contain high densities of sharp edges and vertices. Moreover, hotspots with SERS intensities exceeding those outside the hotspots by several orders of magnitude exist in narrow gaps between almost touching metallic objects.[19, 23-26] An ideal architecture for SERS-based sensing and heterogeneous catalysis would, therefore, comprise dense forest-like layers of overlapping precious metal nanodendrites attached to solid substrates. Dendritic nanostructures grown on solid substrates are often two-dimensional objects extending only in the substrate plane. The precious metal nanodendrites used for SERS-based sensing and heterogeneous catalysis would ideally be three-dimensional (3D) objects that also extend in the direction normal to the substrate plane. Only 3D nanodendrites will have high specific surfaces. Taking into account the extension of the focal volume of confocal laser scanning microscopes typically used for SERS analytics into the direction normal to the substrate surface, 3D precious metal nanodendrites occupying a larger fraction of the focal volume than corresponding 2D objects will help achieve better SERS detection sensitivities.

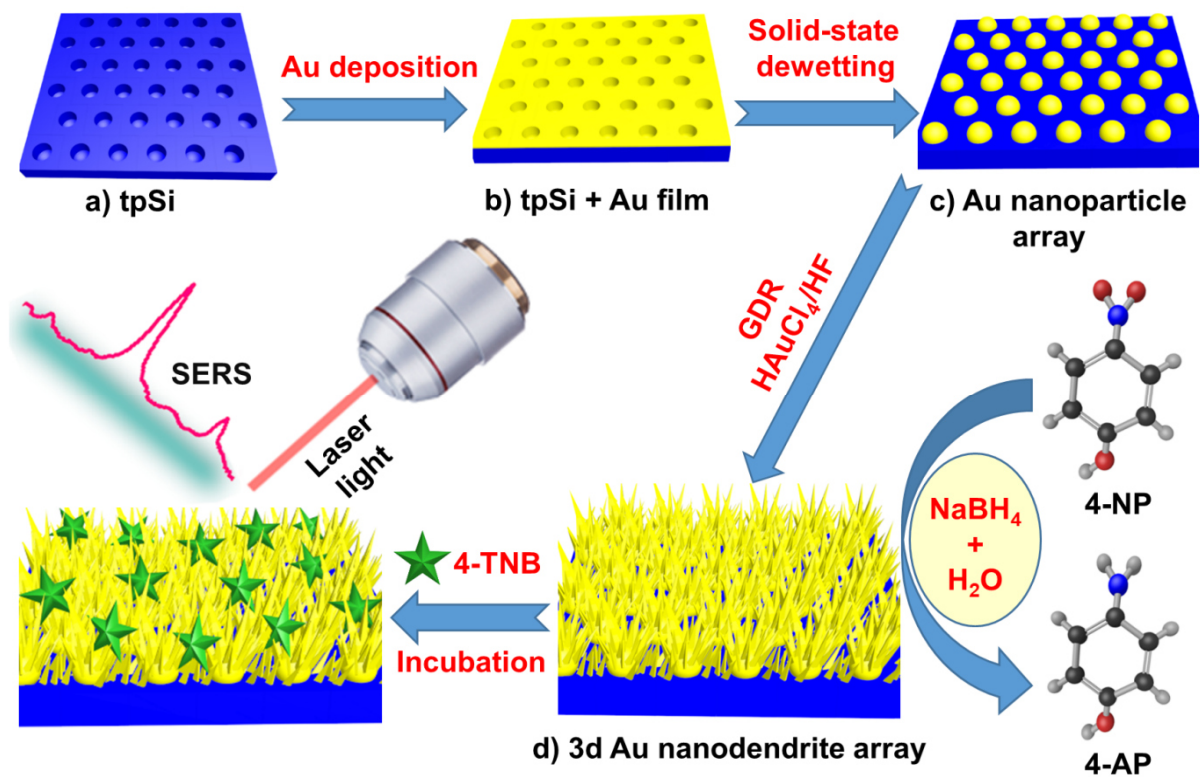

**Figure 1.** Preparation and use of dense layers of overlapping 3D gold nanodendrites attached to silicon wafers. a) Silicon topographically patterned with a regular array of indentations (tpSi; blue) is obtained by capillary microstamping of metal precursors and metal-assisted chemical etching. b) A 35 nm thick gold film (yellow) is deposited on the tpSi. c) Thermal annealing at 900°C converts the 35 nm thick Au film into an array of gold particles, whereby one gold particle forms per tpSi indentation. d) The gold particles act as seeds for the growth of 3D gold nanodendrites by a galvanic displacement reaction (GDR) templated by the gold particle array. The catalytic performance of the dense layers of overlapping 3D gold nanodendrites was tested by the reduction of 4-nitrophenol (4-NP) to 4-aminophenol (4-AP); preconcentration sensing by SERS was tested with the model dye 4-NTB (green).

It has remained challenging to integrate dense 3D precious metal nanodendrite layers in device components suitable for real-life use. Light-induced deposition of gold nanostructures[27] as well as ion implantation[28] have been employed to direct the morphological evolution of the gold nanodendrites themselves. However, it is desirable to initiate the growth of gold nanodendrites at specific sites on a substrate in combination with a certain degree of morphological growth control, aiming at the preparation of dense layers of overlapping 3D gold nanodendrites. In this work, we evaluate a synthetic algorithm yielding dense layers of overlapping 3D gold nanodendrites attached to silicon substrates (Figure 1), in which the growth of the 3D gold nanodendrites starts from gold particles rationally arranged in ordered arrays. We first prepared topographically patterned silicon (tpSi) patterned with arrays of microindentations (Figure 1a) by capillary microstamping.[29] Then, we generated regular arrays of gold particles ~300 nm in diameter by topographically guided solid-state dewetting of thin gold films deposited onto the tpSi. The solid-state dewetting step resulted in the formation of one gold particle per indentation (Figure 1b,c). We used the regularly arranged gold particles as seeds for the templated growth of the 3D gold nanodendrites by galvanic displacement reactions (GDRs). In this way, we obtained dense layers of overlapping 3D gold nanodendrites attached to underlying tpSi substrates (Figure 1d), which showed SERS sensitivities and catalytic performances superior to those of gold dendrites randomly grown on silicon substrates, of silicon substrates covered by thin gold films and of gold particle arrays.

RESULTS AND DISCUSSION

**Gold particle arrays by solid-state dewetting.** Besides electrodeposition[30, 31] and catalytic reduction of precious metal ions[32] GDRs are the most viable access to precious metal dendrites. When a plating solution containing precious metal atoms in a positive oxidation state is brought into contact with an ignoble sacrificial substrate, the precious metal atoms are spontaneously reduced and deposited in the elemental state. The corresponding oxidation of the sacrificial substrate occurs spatially separated from the reduction of the precious metal atoms and results in partial dissolution of the sacrificial substrate. Silver and gold dendrites were obtained by GDRs on sacrificial aluminum,[4, 9] copper,[5, 11, 12] zinc,[2, 13, 33] and nickel[3] substrates. If aqueous plating solutions containing HF and tetrachloroaurate cations are brought into contact with silicon surfaces, gold is deposited by a GDR whereas the silicon is partially oxidized to $[SiF_6]^{2-}$.[34] GDRs on silicon wafers yielded silver and gold dendrites[27, 28, 35, 36] as well as arrays of Au mushrooms.[37] Dense arrays of colloidal-like gold nanostructures decorated with spike-like surface features[8, 11] have been identified as advantageous configurations for SERS substrates. However, GDRs often yield elongated fir branch-like gold microstructures.[2-5, 9, 12, 13, 33] Moreover, the partial dissolution of the sacrificial substrates is hardly controllable. This lack of structural control impedes reliable and reproducible SERS-based analytics and catalytic performance.

3D gold nanodendrite growth by galvanic displacement reactions templated by rationally arranged gold particles is associated with local dissolution of the sacrificial substrate in well-defined interstices not disturbing the generation of the 3D gold nanodendrites. We first prepared tpSi exhibiting regular arrays of indentations with a diameter of ~1.10 µm, a depth of ~200 nm and a nearest-neighbor distance of 1.5 µm (Figure 1a; Figures S1 and S2, Supporting Information) following procedures reported elsewhere.[29] Using topographically patterned polymeric stamps containing continuous nanopore systems for ink supply, we stamped regular hexagonal arrays of submicron silver nitrate dots onto silicon wafers from which the native oxide layer had been removed. Indentations were then formed at the positions of the submicron silver nitrate dots by metal-assisted chemical etching. Apart from the micron-sized indentations tpSi contains surface mesopores as second hierarchical structure level (Figure S2, Supporting Information). In the next step, we deposited 35 nm thick gold films onto the tpSi (Figure 1b). Annealing to 900°C for 2 h transformed the gold films into regular arrays of gold particles, whereby one gold particle was located in each tpSi indentation (Figure 1c; Figure 2). The underlying structure formation mechanism, referred to as solid-state dewetting,[38] involves surface diffusion of metal atoms and clusters at temperatures below the melting point of the corresponding metal driven by capillarity effects. On curved substrates curvature-induced gradients in the chemical potential induce surface diffusion of the metal from convex substrate regions such as protrusions, vertices or ridges to substrate regions with concave curvature, such as hemispherical indentations or grooves. Vice versa, protrusions, vertices or ridges act as diffusion barriers that impede diffusion of the metal out of substrate regions with concave curvature. On substrates containing regular arrays of pits or indentations dewetting may result in the formation of metal particles inside the pits and indentations.[39-44] The structure evolution in the course of solid-state dewetting can well be monitored by varying the annealing temperature. Annealing for 2 h at 400°C results in partial breakup of the gold films; several indentations are, however, still completely covered by a continuous gold layer (Figure S3a, Supporting Information). Annealing for 2 h at 500°C results in the concentration of the Au films within the indentations, whereas the ridges separating the indentations are dewetted (Figure S3b, Supporting Information). After annealing for 2 h at 600°C the beginning of the formation of discrete gold particles inside the indentations is evident (Figure S3c, Supporting Information).

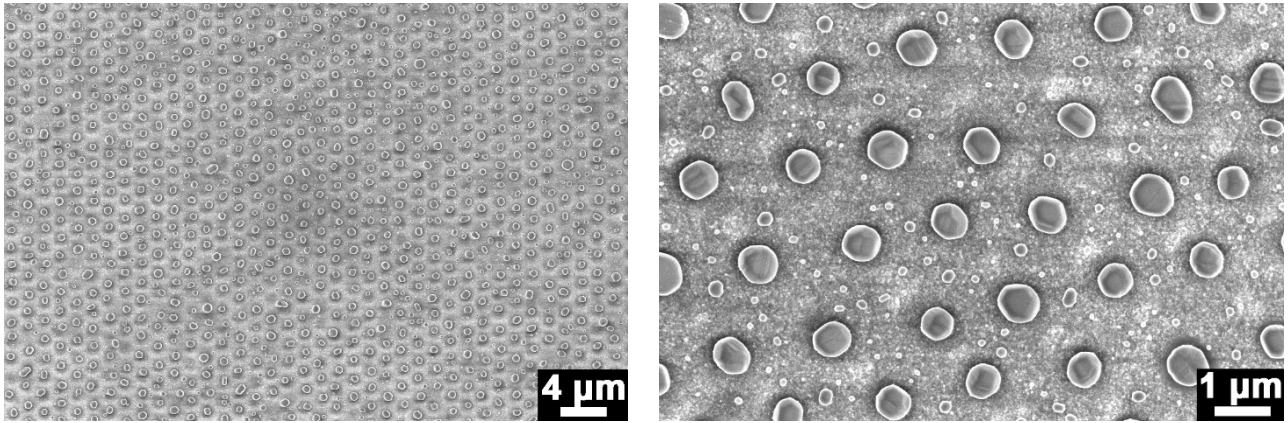

**Figure 2.** a) SEM images of gold-coated tpSi after annealing for 2 h at 900°C. a) Large-field view and b) detail.

Image analysis of scanning electron microscopy (SEM) images (Figure S4, Supporting Information) revealed a decrease in the mean apparent area of the identified discrete gold particles from 0.41 μm$^2$ ± 0.09 μm$^2$ (after annealing at 700°C for 2 h) to 0.33 μm$^2$ ± 0.09 μm$^2$ (after annealing at 800°C for 2 h) to 0.30 μm$^2$ ± 0.06 μm$^2$ (after annealing at 900°C for 2 h) to finally 0.19 μm$^2$ ± 0.03 μm$^2$ after annealing at 1000°C for 2 h. The circularity $C = 4\pi * area/perimeter^2$ is a geometric shape descriptor that quantifies the deviation of an object's contour from that of an ideal circle, for which the circularity equals 1. The circularities of the gold particles increased from 0.80 ± 0.07 (after annealing at 700°C for 2 h) to 0.90 ± 0.04 (after annealing at 800°C for 2 h) and 0.90 ± 0.02 (after annealing at 900°C for 2 h) to finally 0.91 ± 0.02 (after annealing at 1000°C for 2 h). Notable is the decrease in the standard deviations of apparent area and circularity with increasing annealing temperature. It should also be noted that smaller gold particles (a few 10 nm in diameter) were located on the ridges between the indentations of the tpSi, likely pinned at the positions of larger tpSi mesopores (Figure 2b).

**Growth of 3D gold nanodendrite layers by templated galvanic displacement reactions.** The size of the formed gold particles and, therefore, the contact area to the underlying tpSi strongly decreased when the solid-state dewetting temperature was increased from 900°C to 1000°C, possibly resulting in poorer adhesion of the gold particles to the tpSi. Hence, we carried out GDRs on gold-covered tpSi annealed at 900°C for 2 h (Figure 1d) using an aqueous plating solution containing HAuCl$_4$ and HF. Figure 3 shows SEM images of gold particles after different reaction times. Already after 30 s, gold needles characterized by sub-100-nm dimensions started to protrude from the gold particles located inside the indentations of the tpSi (Figure 3a) that further grew with increasing reaction time (Figure 3b). After ~2 min the gold needles exceeded the dimensions of the parent gold particles and began to branch (Figure 3c). After 5 minutes these large gold needles, which were apparently located at the edges of the gold particles, started to develop dendritic features, whereas a second population of smaller sub-100-nm gold needles protruded from smooth surface areas of the gold particles (Figure 3d, see also Figure 4a). After 10 minutes, dendritic and flowerlike structures emanating from the parent gold particles overlapped with their nearest neighbors (Figure 1d, Figure 3e) After 15 minutes densified gold layers characterized by agglomerated platelet-like structures were observed (Figure 3f).

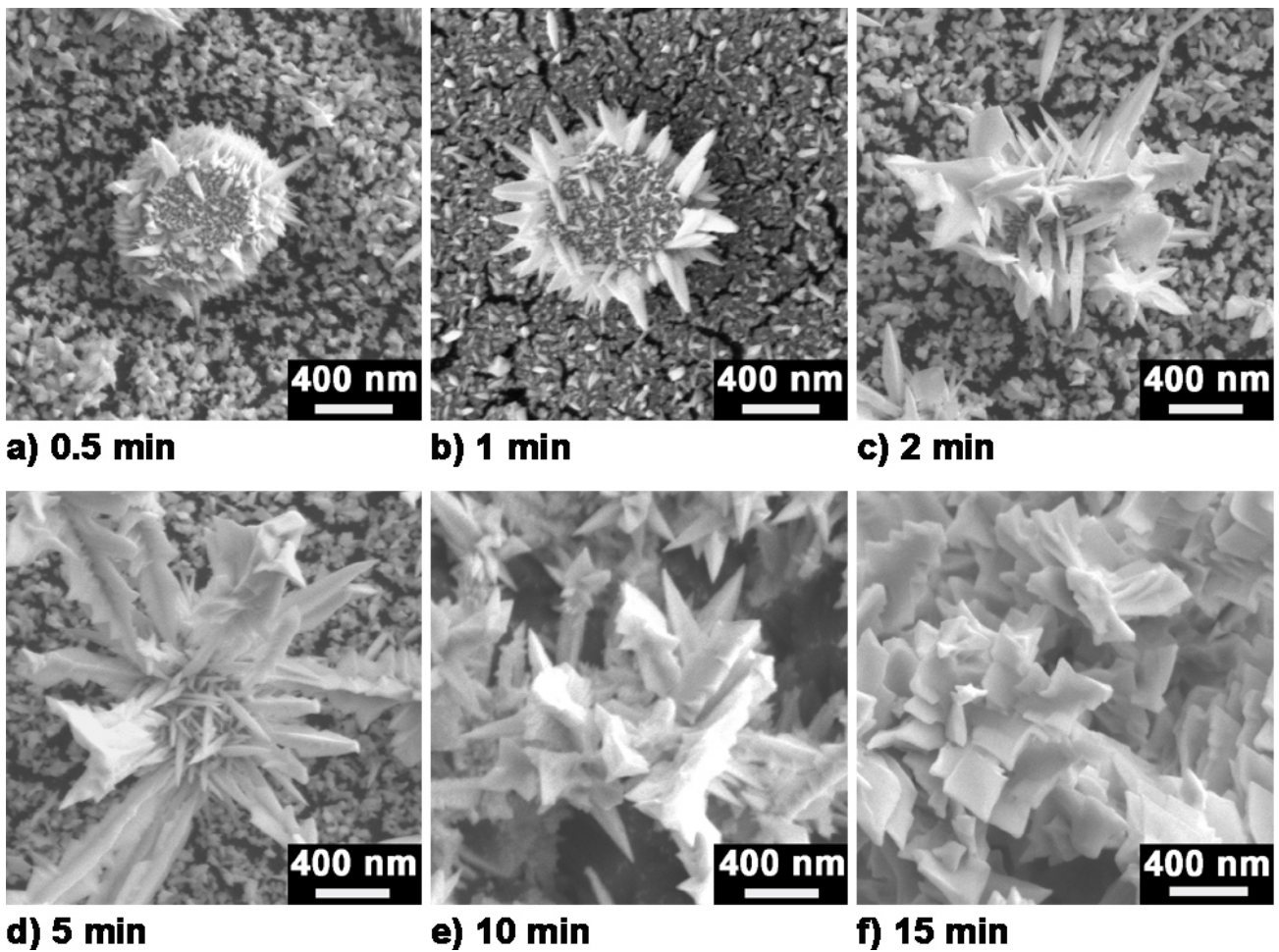

**Figure 3.** Structure evolution for different GDR durations on gold-coated tpSi subjected to solid-state dewetting for 2 h at 900°C.

The dimensions of the gold particles (a few 100 nm) and the spacing between them of 1.5 μm are suitable to ensure that the growth of the individual 3D gold nanodendrites is actually templated by the positioning of the individual gold particles. The transition from separated gold nanodendrites covering only a small portion of the tpSi surface after 5 minutes GDR for to a dense layer of overlapping gold nanodendrites completely covering the tpSi surface after 10 minutes GDR is obvious from a comparison of the lower-magnification SEM images seen in Figure 4a and b. The aggregates of platelet-like gold structures formed after 15 minutes exceeded the nearest-neighbor distance of the gold particles (Figure 4c). Hence, lithographic guidance by the positioning of the parent gold particles faded. Instead, significant amounts of elongated branched gold structures with lengths of several 10 microns, which were irregularly distributed, formed as a second hierarchical structure level (Figure 4d).

Control experiments confirmed that the presence of regularly arranged gold particles acting as seeds during the GDRs is crucial for the generation of dense layers of overlapping 3D gold nanodendrites. Carrying out a GDR for 10 minutes on a smooth silicon wafer coated with a 35 nm thick gold layer (Figure 4e and f; sample "Si + Au" in Figure 5) yielded inhomogeneous substrate coverage with gold structures of various sizes; on the nanoscale, a non-dense layer of polyhedral gold particles formed (Figure 4e). On the macroscale, the same type elongated branched gold structures with sizes of several 10 µm as those seen in Figure 4d were irregularly distributed on the sample surface. Carrying out a GDR for 10 min on tpSi coated with a 35 nm thick gold layer but not subjected to solid-state dewetting yielded non-dense layers of polyhedral gold nanocrystals located on the ridges separating the indentations of the tpSi (Figure S6, Supporting Information; sample "tpSi + Au" in Figure 5).

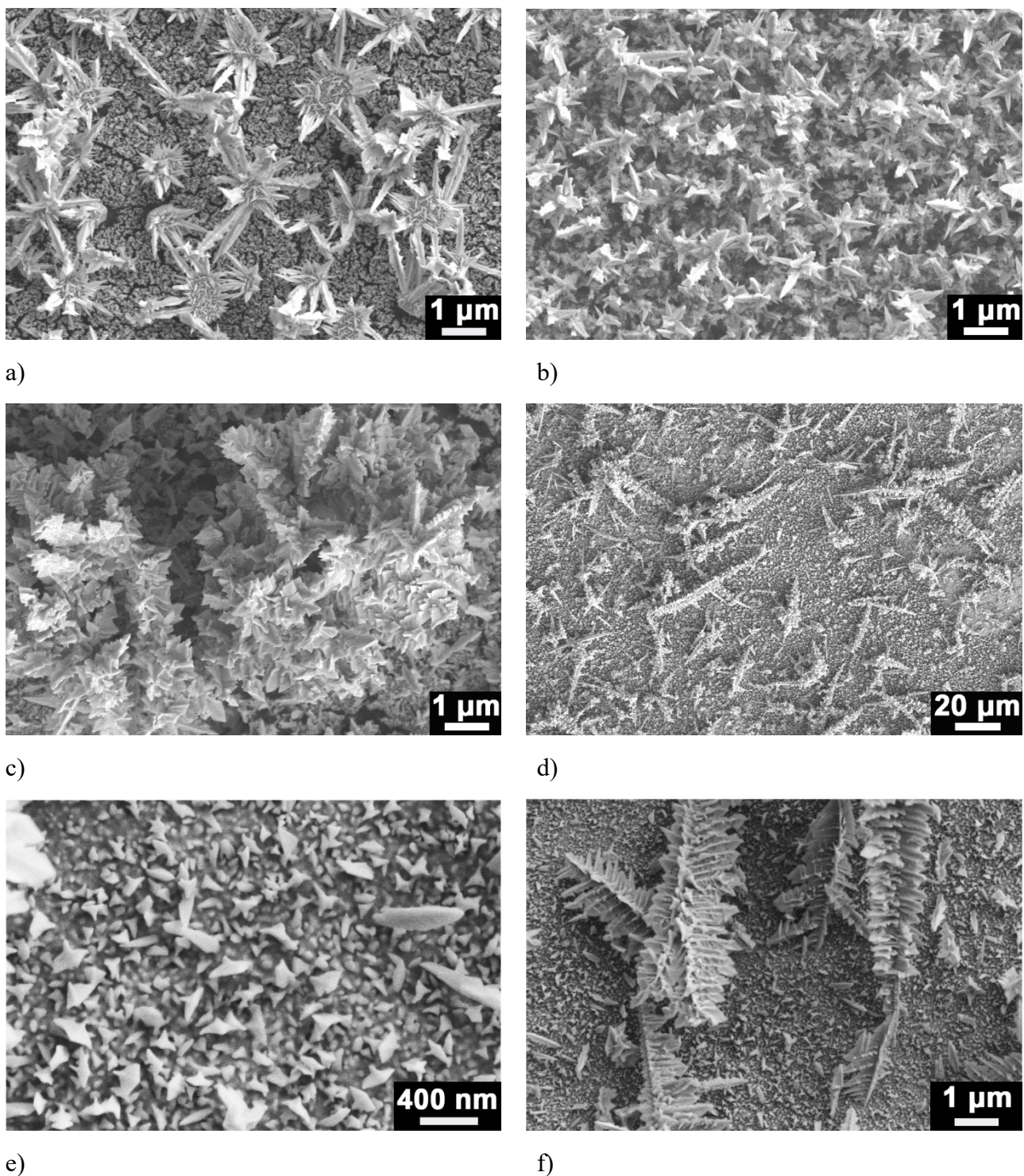

a)

b)

c)

d)

e)

f)

**Figure 4.** Scanning electron microscopy images of GDR products. a) 3D gold nanodendrites on tpSi obtained by solid state dewetting at 900°C for 2 h and GDR for 5 min; b) 3D gold nanodendrites on tpSi obtained by solid state dewetting at 900°C for 2 h and GDR for 10 min; c) detail and d) overview showing the gold structures on tpSi obtained by solid state dewetting at 900°C for 2 h and GDR for 15 min. e) Detail and f) overview of the result of 10 min GDR on a gold-coated Si wafer without topographic patterning (referred to as "Si + Au" in Figure 5).

**SERS performance of 3D gold nanodendrite layers.** Fang and co-authors[8] previously reported that the functionalization of colloidal gold particles with a sea urchin-like multitipped surface via a secondary nucleation process resulted in high SERS enhancement. Further SERS enhancement was achieved by assembling the urchin-like, multitipped gold particles into arrays containing additional hot-spots between adjacent gold particles, whereby the dimensions of the latter where comparable with those of the gold nanodendrites obtained in this work. Here, the two preparative steps boosting SERS performance were integrated into a synthetic algorithm yielding dense arrays of gold nanostructures optimized regarding their SERS performance that are tightly connected to solid substrates, as required for facile recovery. Thus, we evaluated the SERS performance of samples obtained by solid-state dewetting of gold-coated tpSi at 900°C for 2 h followed by GDRs for 0 s, 30 s, 1 min, 2 min, 5 min, 8 min, 10 min, 12 min and 15 min. As reference samples we tested a smooth silicon wafer coated with a 35 nm thick gold layer that was subjected to a GDR for 10 min ("Si + Au", cf. Figure 4e and f), as well as tpSi coated with a 35 nm thick gold layer that was subjected to a GDR for 10 min ("tpSi + Au"; cf. Figure S6, Supporting Information). Sample pieces with areas of 0.25 cm$^2$ were immersed into 1 mL of a 10 $\mu M$ ethanolic solution of 2-nitro-5-thiobenzoate (4-NTB) for 12 h and then washed with ethanol and water. For each sample we measured Raman spectra at 10 different spots. The averaged Raman spectra thus obtained (Figure 5a) show characteristic 4-NTB peaks[45, 46] at ~1340 cm$^{-1}$ (symmetric nitro stretch) and at ~1560 cm$^{-1}$ (aromatic ring mode). Figure 5b displays the relative heights of the 4-NTB Raman peak at 1340 cm$^{-1}$. As compared with the gold particle array on tpSi obtained by solid-state dewetting at 900°C for 2 h without GDR ("0 s") even after short GDR times a doubling of the peak heights occurs, indicating that the formation of needle- and dendrite-like gold structures gives rise to more efficient SERS. The largest peak height is obtained for a GDR duration of 10 min (cf. Figure 3e and 4b), indicating that the dense substrate coverage by touching needle- and dendrite-like gold structures created a high density of SERS hotspots. The peak height exceeds that obtained without GDR (GDR duration of 0 s) by one order of magnitude. Moreover, the Raman peak at 1340 cm$^{-1}$ was 11.3 times higher than for sample "Si + Au" and 12.1 times higher than for sample "tpSi + Au". If the GDR is carried out for more than 10 min the peak heights slightly decrease, indicating deterioration of SERS hotspots as needles, edges and vertices partially buried (Figure 3f; 4c).

The guidance of the growth of the 3D gold nanodendrites by arrays of rationally positioned parent gold particles resulted in reasonably homogeneous SERS response over wide areas. For example, we performed confocal Raman microscopic mapping on 3D gold nanodendrite layers on tpSi obtained by solid-state dewetting for 2 h at 900°C and 10 min GDR within an area of 43×39 µm$^2$ corresponding to overall 1677 Raman spectra. Figure 5c displays the mapping of the integrated Raman intensity of the 4-NTB peak at 1340 cm$^{-1}$. Figure 5d shows the corresponding histogram, i.e., the frequency of integrated peak intensities of the 4-NTB peak at 1340 cm$^{-1}$ falling into specific bins. No local "cold spots" with low integrated peak intensities smaller than 7500 counts were found within the mapped area.

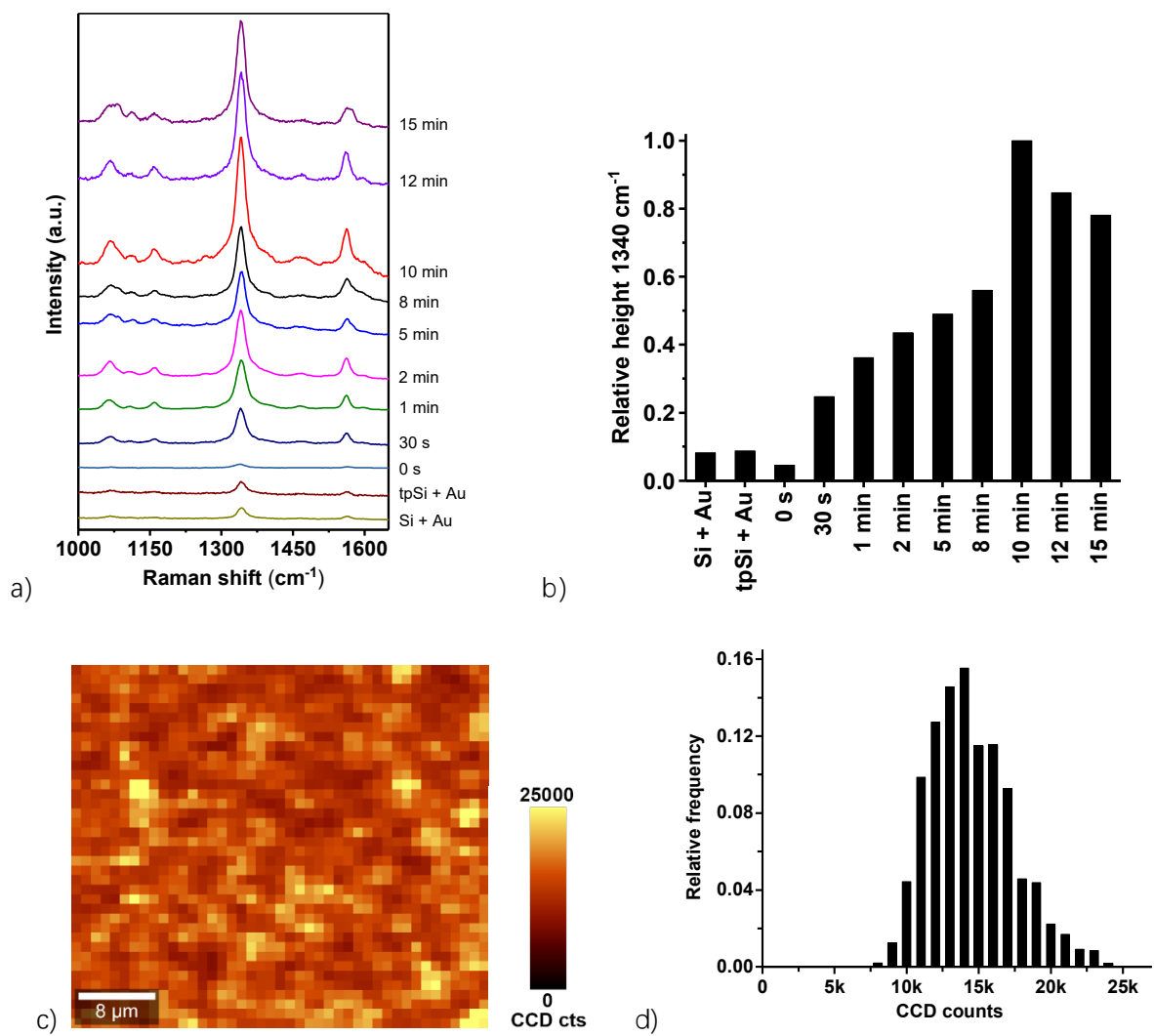

**Figure 5.** a) Raman spectra of GDR products after 4-NTB incubation. Investigated samples include gold particle arrays on tpSi (solid-state dewetting at 900°C for 2 h) after different GDR durations, a smooth gold-coated Si wafer subjected to 10 min GDR ("Si + Au"; Figure 4e,f) as well as gold-coated tpSi subjected to 10 min GDR without solid-state dewetting ("tpSi + Au"; Supporting Figure S6). Each spectrum displayed in panel a) is the average of 10 Raman spectra measured at different positions. b) Relative heights of the 4-NTB Raman peaks at 1340 cm$^{-1}$ extracted from the Raman spectra displayed in panel a). c) Mapping of the integrated Raman intensities of the 4-NTB peak at ∼1340 cm$^{-1}$ on a 3D gold nanodendrite layer prepared by solid-state dewetting for 2 h at 900°C and 10 min GDR; 1677 spectra were captured in an area of 43×39 μm$^2$. d) Histogram to panel c) displaying the relative frequencies of occurrence of the integrated Raman intensities of the 4-NTB peak at 1340 cm$^{-1}$.

**Catalytic performance of 3D gold nanodendrite layers.** The reduction of 4-nitrophenol (4-NP) to 4-aminophenol (4-AP) in the presence of sodium borohydride (NaBH$_4$) in aqueous solutions is a well-established pseudo-first-order model reaction for the evaluation of the catalytic performance of metal nanostructures.[47] 4-NP in aqueous solution shows an absorption band at 317 nm. The conversion of colorless 4-NP to yellow 4-nitrophenolate triggered by the addition of NaBH$_4$ is accompanied by the appearance of a new peak at 400 nm related to changes in the electronic structure (Figure S7, Supporting Information). When monitored by UV-vis spectroscopy, the conversion of 4-NP to 4-AP is apparent from the successive decrease in the detected absorbance at 400 nm. Figure 6a shows the absorption spectra for 4-NP reduction in the presence of a 3D gold nanodendrite layer on tpSi with an area of 0.25 cm$^2$ obtained by solid-state dewetting for 2 h at 900°C and subsequent GDR for 10 min corresponding to Figures 3e and 4b. We fitted -ln[$A_t/A_0$], where $A_t$ is the absorbance at reaction time $t$ and $A_0$ the absorbance at $t = 0$ s, as function of $t$ using a first-order rate law. The slope of -ln[$A_t/A_0$]($t$) then corresponds to the pseudo-first-order rate constant $k$. In the presence of a 3D gold nanodendrite layer on tpSi prepared by solid-state dewetting for 2 h at 900°C and subsequent GDR for 10 min $k$ amounted to ~0.09 min$^{-1}$ (Figure 6b, black squares). As obvious from SEM investigations (Figure S9, Supporting Information), the 3D gold nanodendrites were still intact after use in the reduction of 4-NP. The $k$ value obtained in the presence of sample "tpSi + Au" – gold-coated tpSi after 10 min GDR but without solid-state dewetting (Figure S8a; red circles in Figure 6b) – amounted to only ~0.02 min$^{-1}$. The $k$ values obtained in the presence of smooth silicon coated with a 35 nm thick gold layer without further treatment (Figure S8b; blue up-triangles in Figure S6b) and in the presence of gold-coated tpSi subjected to solid-state dewetting for 2 h at 900°C without GDR (Figure S8c; orange down-triangles in Figure 6b) were even smaller and amounted to ~0.009 min$^{-1}$ and ~0.002 min$^{-1}$, respectively (all above-mentioned samples had an area of 0.25 cm$^2$). In the presence of a 3D gold nanodendrite layer on tpSi prepared by solid-state dewetting for 2 h at 900°C and subsequent GDR for 10 min ~92 % of the 4-NP was converted at $t = 24$ min, in the presence of gold-coated tpSi after 10 min GDR without solid-state dewetting ~40 %, in the presence of smooth silicon coated with a 35 nm thick gold layer ~21 % and in the presence of gold-coated tpSi subjected to solid-state dewetting for 2 h at 900°C without GDR ~5 %.

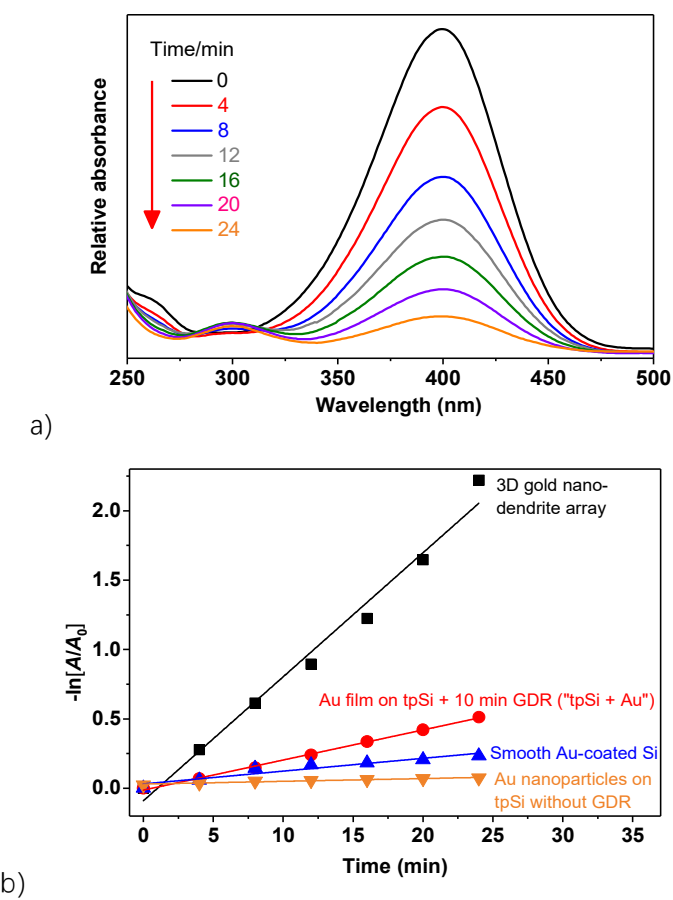

**Figure 6.** a) UV-vis absorption spectra showing the absorption band of 4-NP in the presence of $NaBH_4$ at 400 nm during 4-NP reduction catalyzed by a 3D gold nanodendrite array on tpSi obtained by solid-state dewetting for 2 h at 900°C and 10 min GDR corresponding to Figures 3e and 4b. b) Plots of $-\ln[A_t/A_0]$ versus reaction time $t$ for the reduction of 4-NP in the presence of $NaBH_4$. $A_t$ is the absorbance of 4-NP at 400 nm at reaction time $t$; $A_0$ is the absorbance of 4-NP at 400 nm at $t = 0$ s. Black squares: 3D gold nanodendrite array on tpSi obtained by solid-state dewetting for 2 h at 900°C and 10 min GDR; red circles: gold-coated tpSi after 10 min GDR without solid-state dewetting (sample "tpSi + Au"); blue up-triangles: smooth silicon wafer coated with a 35 nm thick gold layer; orange down-triangles: gold particle array on tpSi without GDR obtained by solid-state dewetting for 2 h at 900°C. The fits were obtained by linear regression.

CONCLUSIONS

We have evaluated the use of rationally positioned gold particles as means to direct the generation of dense, homogeneous layers of overlapping three-dimensional gold nanodendrites on silicon wafers for applications in the fields of SERS-based sensing and heterogeneous catalysis. The first preparation step comprised the generation of topographically patterned silicon; ordered arrays of metal precursor microdots were deposited on smooth silicon wafers by capillary microstamping. Subsequent metal-assisted chemical etching yielded indentations at the positions of the metal precursor microdots. Thin gold films deposited on topographically pattered silicon were subjected to solid-state dewetting yielding one gold particle per indentation. In subsequently performed galvanic displacement reactions the regularly arranged gold particles with diameters of ~300 nm and a nearest neighbor distance of ~1.5 μm acted as seeds for the three-dimensional growth of gold nanodendrites. Size and spacing of the parent gold particles resulted in the initiation of the growth of individual 3D gold nanodendrites at individual gold particles. The emergence of dendritic structure features at the edges and vertices of the gold particles led to the formation of 3D gold nanodendrites rather than of 2D gold nanodendrites parallel to the substrate plane. Partial dissolution of the underlying topographically patterned silicon took place at the interstices between the growing gold nanodendrites. Templating the growth of the 3D gold nanodendrites by rationally positioned parent gold particles is key to the growth of dense layers of overlapping 3D gold nanodendrites with reasonable homogeneity over large areas. By optimizing the GDR duration, 3D gold nanodendrite layers characterized by large specific gold surfaces as well as by abundance of sharp, nearly touching gold edges and vertices were obtained. The homogeneous layers of overlapping three-dimensional gold nanodendrites optimized in this way exhibited performances in heterogeneous catalysis and SERS-based preconcentration sensing superior to configurations obtained without rational prepositioning of the gold nanodendrite growth sites.

## MATERIALS AND METHODS

**Materials.** $p$-type (100)-oriented silicon wafers with a resistivity of 1−3 Ω cm were provided by Siegert Wafer. Gold (III) chloride hydrate ($HAuCl_4 \cdot xH_2O$, 99.995%), sodium borohydride ($NaBH_4$, 98%), 4-nitrophenol (4-NP) and absolute ethanol were purchased from Sigma-Aldrich. Hydrofluoric acid (HF, 48%) was obtained from Merck (Darmstadt, Germany). 5,5′-dithiobis (2-nitrobenzoic acid) (DTNB) was purchased from Alfa Aesar. Topographically patterned silicon (tpSi) was prepared following procedures reported elsewhere,[29] the difference being that spongy polystyrene-block-poly(2-vinylpyridine) (PS-$b$-P2VP) stamps topographically patterned with hexagonal arrays of rod-like contact elements having hemispherical tips (height 950 nm; base diameter 1.0 μm) were used. The spongy PS-$b$-P2VP stamps were generated by molding PS-$b$-P2VP against macroporous silicon templates (provided by SmartMembranes, Halle, Germany) containing correspondingly shaped macropores (diameter of macropore mouths 800 nm; depth of macropores 1.0 μm). Capillary microstamping thus yielded circular $AgNO_3$ dots with a diameter of

~1.15 μm and a height of ~6.0 nm arranged in hexagonal arrays with a lattice constant of 1.5 μm. Residual silver and silver nitrate after the metal-assisted chemical etching step was removed by treatment in 30 mL $HNO_3/H_2O$ (v/v = 1:1) solution.

**Solid-state dewetting.** Smooth Si wafer pieces and tpSi pieces with areas of 0.5×0.5 $cm^2$ were coated with a 35 nm thick gold layer using a Balzers BAE 120 evaporator following protocols reported elsewhere.[48] For this purpose, the tpSi and smooth Si wafer pieces were glued on glass slides with double-sided adhesive tape and placed in the vacuum chamber above the gold source at a distance of 25 cm. The chamber pressure was set to $10^{-4}$ mbar, the current was set to 4.20 A, and the baffle was removed for gold deposition. The thickness of the obtained gold was estimated by weighting glass slides with known surface area prior to and after gold deposition using a quartz crystal microbalance GAMRY eQCM 10M. Solid-state dewetting was carried out in a tube furnace by heating gold-coated tpSi at a rate of 10 K/min to the target temperature at which the gold-coated tpSi was kept for 2 h. Then, the samples were cooled to room temperature at the natural cooling rate of the tube furnace.

**Galvanic displacement reactions (GDRs).** Prior to the GDRs, the samples were rinsed with diluted $HCl_{(aq)}$ (2 mol/L) for 4 min, treated with $O_2$ plasma at 100 W for 2 min using a plasma etcher Femto (Diener Electronics) to remove contaminations and then immersed in a 5% aqueous HF solution for 2 min to remove native silica. Then, the samples were immediately immersed in an aqueous plating solution containing 20 mL $HAuCl_4 \cdot xH_2O$ (20 mmol/L) and 18 mL HF (48%) at room temperature.

**Characterization**. Scanning electron microscopy was carried out on a Zeiss AURIGA microscope operated at an accelerating voltage of 7 kV using a secondary electron (SE) detector to image the 3D gold nanodendrites and an in-lens detector to characterize the gold particle arrays obtained by solid-state dewetting. Image analysis of the SEM images was carried out with the software ImageJ using the freehand selection tool.

**Raman measurements.** 4-NTB in ethanolic solution was obtained by cleavage of the disulfide bond of DTNB with $NaBH_4$. All samples characterized by Raman spectroscopy had an area of 0.25 $cm^2$. Each sample piece was immersed into 1.0 mL of a solution of 4-NTB in ethanol ($10^{-5}$ mol/L) for 12 h under shaking. Then, the samples were washed with ethanol and water for three times and dried under $N_2$ flow. All Raman measurements were carried out with a Raman microscope WITec

Alpha 300R (30 cm focal length and 600 grooves per mm grating spectrometer) equipped with an EM-CCD (Andor Newton DU970N–BV-353) using the 632.8 nm line of a He-Ne laser with a power of 0.4 mW at the sample. The Raman spectra were collected under 40x magnification (NA = 0.6) using an air objective with glass correction. Figure 4a shows the averages of 10 Raman spectra measured at different positions on the respective samples with an integration time of 0.5 s. For the mapping displayed in Figure 4c and d Raman spectra with an integration time of 0.2 s were acquired. The Raman spectra were analyzed and processed with the software Project FOUR.

**Catalytic reduction of 4-nitrophenol.** Aqueous solutions of 4-NP (1.33 mL, 0.1 mmol/L) and solutions of $NaBH_4$ (1 mL, 0.1 mol/L) cooled to 0°C were mixed with 0.67 mL deionized water in standard quartz cuvettes with a path length of 1.0 cm. Then, sample pieces with areas of 0.25 cm$^2$ were immersed into the mixture. For the measurements of the UV-vis absorption spectra with a duration of 2 min the reaction mixtures were separated from the sample pieces by transfer into a second identical cuvette. The UV-vis absorption spectra were acquired with a UV-Vis-NIR spectrophotometer Cary 6000i in the wavelength range from 250 nm to 500 nm at ambient temperature. For the reaction time $t$ only the time the reaction mixtures were in contact with the sample pieces were considered. We fitted $-\ln[A_t/A_0]$ as function of $t$ assuming a first-order rate law by linear regression with the software Origin 2017.

## Conflicts of interest

There are no conflicts to declare.

## Acknowledgements

The authors thank the European Research Council (ERC-CoG- 2014, project 646742 INCANA) for funding.

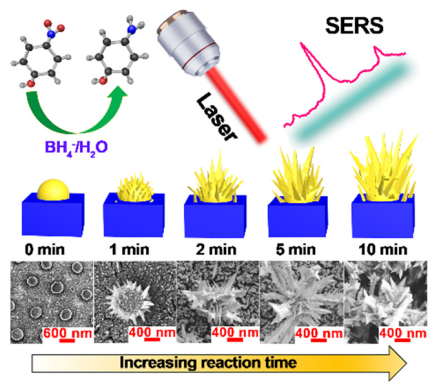

TOC Figure

**Dense layers of overlapping three-dimensional gold nanodendrites** obtained by lithographically guided gold nanodendrite growth were evaluated for SERS-based preconcentration sensing and heterogeneous catalysis.